\documentclass[aps,floatfix]{revtex4}
\usepackage{amsfonts}%
\usepackage{amsmath}%
\usepackage{amssymb}%
\usepackage{graphicx}
\providecommand{\U}[1]{\protect\rule{.1in}{.1in}}

\newcommand{\half}{{\textstyle \frac{1}{2}}}
\newcommand{\quarter}{{\textstyle \frac{1}{4}}}

\newcommand{\bel}[1]{\begin{equation}\label{#1}}
\newcommand{\bal}[1]{\begin{eqnarray}\label{#1}}
\newcommand{\ee}{\end{equation}}
\newcommand{\ea}{\end{eqnarray}}
\newcommand{\equ}[1]{~Eq.(\ref{#1})}

\newcommand{\zb}{\bar{z}}

\newcommand{\bmath}{\begin{displaymath}}
\newcommand{\emath}{\end{displaymath}}
\newcommand{\bite}{\begin{itemize}}
\newcommand{\eite}{\end{itemize}}

\newcommand{\eps}{\varepsilon}
\newcommand{\drop}[1]{}

\begin{document}
\preprint{HEP-TH/1000708}
\title[SelfCasimir]{Semiclassical Estimates of Electromagnetic Casimir Self-Energies of Spherical and Cylindrical Metallic Shells}
\author{Martin Schaden}
\affiliation{Department of Physics, Rutgers University, 101 Warren Street, Newark NJ 07102}

\begin{abstract}
\noindent The leading semiclassical estimates of the electromagnetic Casimir stresses on a spherical and a cylindrical metallic shell are within 1\% of the field theoretical values. The electromagnetic Casimir energy for both geometries is given by two decoupled massless scalars that satisfy conformally covariant boundary conditions.  Surface contributions vanish for smooth metallic boundaries and the finite electromagnetic Casimir energy in leading semiclassical approximation is due to quadratic fluctuations about periodic rays in the interior of the cavity only. Semiclassically the non-vanishing Casimir energy of a metallic cylindrical shell is almost entirely due to Fresnel diffraction. \vspace{1pc}
\end{abstract}





\startpage{101}
\endpage{120}
\maketitle

\section{Introduction}
Casimir obtained his now famous attractive force between two neutral metallic
plates\cite{Casimir48} by considering the boundary conditions these impose on the electromagnetic field. Half a
century later, his prediction was verified experimentally\cite{Experiments99} to better than 1\%.

Twenty years after Casimir's work, Boyer calculated the zero-point energy of an
ideal conducting spherical shell\cite{Boyer68}. Contrary to intuition derived from the attraction between two
parallel plates, the sphere tends to expand. Boyer's result has since been improved in accuracy and
verified using a number of field theoretic methods\cite{Davies72,BD78,Milton78,EKK00} -- even though there
may be little hope of observing this effect experimentally in the near future\cite{Barton04}.

Since field theoretic methods require explicit or implicit knowledge of cavity \emph{frequencies}, they have
predominantly been used to obtain the Casimir energies of classically \emph{integrable}
systems. Thus, in addition to a spherical cavity, the electromagnetic Casimir energies of dielectric
slabs\cite{Lifshitz56,Milloni93,Milton04}, metallic parallelepipeds\cite{Lukosz71,DC76,AW83} and long
cylinders\cite{DRM81,GR98,MNN99,LNB99,NP00}  have been computed in this manner.

However, most systems are not integrable and often cannot even be approximated by integrable systems. It thus is
desirable to develop reliable methods for estimating Casimir energies of classically non-integrable and even
chaotic systems. Balian, Bloch and Duplantier calculate Casimir energies using a multiple scattering
approximation to the Green's function\cite{BB70,BD04}. This approach does not require knowledge of the quantum
mechanical spectrum and the geometric expansion in principle is exact for sufficiently smooth and ideally
metallic cavities. However, except for some integrable systems, it in practice is often difficult to carry the multiple scattering expansion
beyond its first few terms. The relative importance of terms in this expansion also is difficult to assess a priori.  In\cite{MS98} a semiclassical method was proposed to estimate (finite) Casimir energies. It is based on Gutzwiller's trace formula\cite{Gutzwiller90} for the response function and is suitable for Casimir energies of hyperbolic and chaotic systems\cite{MS98,SS05,BMW06} with isolated classical periodic
orbits.

Although not exact in general, the semiclassical approximation associates the finite (Casimir) part of
the vacuum energy with optical properties of the system. It captures aspects of Casimir energies that have been
puzzling for some time\cite{Schaden06}.  Path integral methods\cite{Schubert02,GLM03,GK05,EHGK01,Schaden09a} in principle
allow one to obtain Casimir forces to arbitrary precision. Due to unresolved
renormalization problems, these methods have not yet been used to study the self-stress on cavities. We here use semiclassical methods adapted to classically integrable systems to estimate and analyze the Casimir self-stress of a spherical and a cylindrical shell.

The simplicity, transparency and surprising accuracy of this approximation is first demonstrated on
Boyer's problem\cite{Boyer68,Davies72,BD78,Milton78,EKK00}, the electromagnetic
Casimir energy of a spherical cavity with an (ideal) metallic boundary.
The semiclassical analysis of this problem is an order of magnitude simpler than any given previously and the positive sign is related to caustic surfaces of second order.
In sect.~4 the semiclassical estimate for the electromagnetic Casimir self-energy of a perfectly
conducting cylindrical shell is reexamined by converting the sum of WKB-estimates for the eigenfrequencies to the dual sum over periodic orbits. It was previously\cite{MSSS03,Schaden06b} found that the Casimir self-energy of a cylindrical metallic shell vanishes to leading semiclassical order. It in fact vanishes only because the upper bound of a particular Fresnel integral is ignored in this approximation. Demanding this physical bound in the longitudinal momentum fraction, the semiclassical estimate also is rather accurate for the self-stress of a metallic cylindrical shell and within $\quarter$\% of its field-theoretic value\cite{DRM81}. The sign of the self-stress is again determined by the presence of caustics and the associated Maslov-Keller indices\cite{Keller58,Maslov72}. However, contrary to the spherical case, this self-stress semiclassically is primarily due to Fresnel diffraction effects. A discussion of the results and a critical assessment of difficulties remaining for a semiclassical interpretation of some Casimir self-energies concludes this article.

\section{The Dual Picture: Electromagnetic Casimir Energies of Integrable Systems and Periodic Rays}
Integrable systems may be semiclassically quantized in terms of periodic paths on invariant
tori\cite{Einstein17} -- in much the same manner as Bohr first quantized the hydrogen atom. Although in general
not an exact transformation, classical periodic orbits on the invariant tori are \emph{dual} to the mode
frequencies in a semiclassical sense. Applying Poisson's summation formula, the semiclassical Casimir energy
(SCE) due to a massless scalar may be written in terms of classical periodic
orbits\cite{BT76,Gutzwiller90,Schaden06,Brackbook},
\bal{dualtrafo}
{\cal E}_c &=& {\frac1 2}\sum_{\bf n} \hbar\omega_{\bf n}\ - \ {\rm UV~subtractions} \nonumber\\
&\sim & \ \frac{ 1}{ 2 \hbar^d}{\sum_{\bf m}}^\prime e^{-\frac{i \pi}{2}\beta_{\bf m}} \int_{sp} {\bf
dI}\, H({\bf I})\, e^{2\pi i\, {\bf m}\cdot{\bf I}/\hbar}\ .\nonumber\\
\ea
The components of the $d$-dimensional vector ${\bf I}$ in\equ{dualtrafo} are the actions of a set of properly
normalized action-angle variables that describe the integrable system. The exponent of the integrand
in\equ{dualtrafo} is the classical action (in units of $\hbar$) of a periodic orbit that winds $m_i$ times about
the $i$-th cycle of the invariant torus.  $H({\bf I})$ is the associated classical energy and $\beta_{\bf m}$ is
the Keller-Maslov index\cite{Keller58,Maslov72} of a class of periodic orbits identified by ${\bf m}$. The
latter is a topological quantity that does not depend on the actions ${\bf I}$. To leading semiclassical order,
the (primed) sum extends only over those sectors ${\bf m}$ with non-trivial stationary points (the classical periodic paths of finite action) (see
below). The correspondence in \equ{dualtrafo} can only be argued semiclassically\cite{Gutzwiller90,BT76,Brackbook} and the
integrals on the RHS should therefore be evaluated in stationary phase approximation $(sp)$ only.

The semiclassical spectrum of a massless scalar is exact for a number of manifolds without
boundary\cite{Dowker71} and the definition of the semiclassical Casimir energy (SCE) by the RHS of \equ{dualtrafo} coincides with the Casimir
energy of zeta-function regularization in these cases. It also is exact for massless scalar fields satisfying
periodic-, Neumann- or Dirichlet- boundary conditions on parallelepipeds\cite{Lukosz71,AW83,Schaden06} as well
as for some tessellations of spheres\cite{Schaden06,CD93,Dowker05}. To physically interpret the finite SCE of a system, one has to consider the implicit subtractions in the spectral density\cite{Barton01,BD04,Schaden06}.

Semiclassical contributions to the spectral density arise due to small fluctuations about closed classical paths. These either are contractible to a point or not. in this dual picture local UV divergences are associated with fluctuations about arbitrary short contractible classical paths.  If all local UV divergences can be subtracted unambiguously\cite{CD79,Barton01,Schaden06}, the dependence of the remaining finite
Casimir energy on macroscopic deformations of the system arises from fluctuations about classical closed paths of finite classical action only.

In disjoint systems, the Casimir interaction at small separations usually is dominated by fluctuations about periodic orbits\cite{Schaden06}. If there are no stationary classical periodic orbits, the leading interaction typically is due to periodic orbits of extremal- rather than stationary- length\cite{SS04}. These correspond to diffractive effects that lead to relatively weak but sometimes rather interesting interactions. Examples of systems without stationary periodic orbits include the Casimir pendulum of\cite{SJ05}, perpendicular plates\cite{GK05}, a wedge above a plate and, most recently, a rotational ellipsoid above a plate with a hole\cite{LMRRJ10}. Diffractive contributions also become important when the separation between two systems is comparable to or larger than the smaller system. The original Casimir-Polder interaction between two atoms\cite{CasimirPolder48} or between an atom and a metallic plate\cite{Spruch} fall in this category.

But periodic (stationary or extremal) orbits need not dominate the Casimir energy when some non-periodic closed orbits of finite action are much shorter than any periodic ones. The Casimir force due to a massless scalar field satisfying Dirichlet boundary conditions on the plate of a hemispherical Casimir piston\cite{Schaden09} is an example. Periodic orbits similarly may not give the largest contribution to the self-stress of a cavity for scalar fields satisfying Dirichlet or Neumann conditions. For sufficiently smooth boundaries like the spherical and cylindrical (toroidal) shell we are interested in, only arbitrary short classical paths that reflect just once off the boundary potentially can lead to UV-divergences. The divergence generally will depend on the local curvature of the boundary\cite{CD79}. Even if this divergence is not logarithmic and can be subtracted unambiguously, the finite remainder of such short closed orbits can contribute significantly to the Casimir energy\cite{Schaden09}.

 The following dimensional argument suggests that in a dimensionless regularization scheme the local ultraviolet divergence due to scalar fields satisfying Dirichlet or Neumann conditions on both sides of a smooth and \emph{infinitesimally thin} even-dimensional surface in fact vanishes: barring other scales, the local contribution from a small $(d-1)$-dimensional surface element $dA$ to
the divergence for dimensional reasons is of the form  $\hbar c f_\eps(R_i/R_j)\,dA/R^d $, where $R$ is the principal (local)
radius of curvature of the surface and $f_\eps$ is a dimensionless function of the (dimensionless) regularization parameter $\eps$ and of ratios of the local curvatures only. The interior and exterior radii of curvature at the same point on the infinitesimally thin surface are of equal magnitude but of opposite sign. Local divergent surface contributions from the interior and exterior of the infinitesimally thin surface [with the same scaleless boundary conditions on \emph{both} of its sides] thus cancel precisely for odd dimensions $d$. For a spherical surface this cancelation has been explicitly observed in\cite{BM94}. In $d=3$, finite Casimir energies have also been calculated for scalar
fields and an infinitesimally thin cylindrical shell\cite{GR98,Saharian00,ST06}. The argument above suggests
that surface divergences in fact cancel locally for any \emph{infinitesimally thin} (and
sufficiently smooth) even-dimensional boundary, regardless of its shape or whether Dirichlet or Neumann conditions are imposed.

However, for scalar fields this cancelation occurs only for vanishingly thin even-dimensional smooth boundaries and in a dimensionless regularization scheme. In a physical cutoff scheme this only implies the absence of logarithmic divergences, and surface divergences do occur even for ideal boundaries. They can be unambiguously subtracted\cite{BKV99,GJKQSW02,Milton02,PMK07} or, equivalently, absorbed in parameters describing physical properties of the surface. Although the subtractions are not ambiguous for even-dimensional ideal interfaces, the remaining finite contribution to the Casimir energy due to closed non-periodic classical paths could be significant and \equ{dualtrafo} in this case can give an inadequate estimate of the Casimir self-stress due to scalar fields [see \cite{Brackbook,Schaden09} and sect.-4 for some formally sub-leading semi-classical surface contributions that have been omitted in\equ{dualtrafo}.]

Fortunately the finite \emph{electromagnetic} self-stress of a smooth and infinitesimally thin perfectly conducting boundary \emph{is} predominantly due to the stationary points of the integrand in \equ{dualtrafo}. The electromagnetic Casimir self-energy of a closed, smooth and perfectly metallic shell may be decomposed into the contributions from two massless scalar fields -- one satisfying Dirichlet's, the other Neumann's boundary condition\footnote{For a spherical shell the usual Robin- and Dirichlet- conditions on solutions  $B_{\ell}(x)$ of the spherical Bessel equation are dictated\cite{Gilkey} by the conformal covariance of free electromagnetic fields. They may be thought of as simple Neumann- and Dirichlet- boundary conditions for radial functions $\phi_\ell(x)=x B_{\ell}(x)$  of vanishing conformal dimension that are solutions of  $(\partial^2_x+1-\ell(\ell+1)/x^2)\phi_\ell(x) =0$.} on the surface\cite{BD04}.  Semiclassically, these boundary conditions are enforced by a phase lag of $\pi$(0) for the Dirichlet(Neumann) scalar at each specular reflection of the classical path. Since both scalar fields satisfy the same wave equation, only fluctuations about classical trajectories with an \emph{even} number of reflections contribute to the electromagnetic spectral density of an ideal metallic cavity\cite{BD04}. Potentially divergent contributions of the Neumann- and Dirichlet- scalar that arise from closed contractible classical paths that reflect just once off the metallic boundary in this case cancel each other \emph{exactly}. The length of any closed classical ray with an \emph{even} number of reflections off a sufficiently smooth cavity (without sharp corners) thus is bounded below by geometrical characteristics of the cavity -- such as its minimal radius of curvature. The closed classical rays of minimal length with an even number of reflections off such cavities are periodic and correspond to either a stationary point or an extremum of the classical action. For the cylindrical- and spherical- shells we will be considering, extremal periodic rays creep about the exterior of the cavity a number of times. They lead to rather small diffractive corrections\cite{Keller62,SS04,Brackbook} that will be ignored. The stationary points of the action in\equ{dualtrafo} give the main contribution in the electromagnetic case. They correspond to periodic trajectories inside the cavity characterized by their winding number and  (even) number of reflections such as those shown in Fig.~1a.

The previous argument applies equally well to the contribution of any pair of decoupled scalar fields satisfying Dirichlet and Neumann boundary conditions and it may not be apparent why the SCE should give a particularly good approximation for the self-stress of a metallic shell due to the \emph{electromagnetic} field. The exact field theoretic Casimir stress caused by two massless scalar fields solving the same Helmholtz equation as the transverse electromagnetic fields but satisfying Neumann- (instead of Robin-) and Dirichlet- boundary conditions on the spherical shell not only differs in sign but also is an order of magnitude larger than the electromagnetic one\cite{BM94,NP97,GR98}: $-0.220967\dots\hbar c/R$ instead of the $+0.04617\dots \hbar c/R$ in the electromagnetic case\cite{Boyer68,Milton78}.  The reason for this difference is that the Neumann boundary condition on a spherical shell is not conformally covariant\cite{Gilkey} for a scalar satisfying the Helmholtz equation whereas the Robin boundary condition is\footnote{$\partial_r (r\phi(r))=0$ is conformally covariant if $\phi$ has conformal mass dimension $1$.}.  On an intrinsically flat boundary like a cylindrical shell, the electromagnetic Casimir energy of a metallic shell indeed decomposes into contributions from two scalar fields satisfying Dirichlet and Neumann boundary conditions\cite{GR98}. The classical action of a massless scalar particle is conformally invariant and the semi-classical approximation indeed reproduces\cite{Schaden06} the Casimir energy of a conformal scalar field on curved manifolds without boundary such as that of a three-dimensional sphere $S_3$. Since specular reflection and phase lag do not depend on the curvature of a surface, we conjecture that for a smooth but curved boundary the SCE approximates the Casimir energy of scalars satisfying \emph{conformally covariant} Dirichlet and Neumann conditions on the boundary. For boundaries with non-vanishing curvature, the latter correspond to Robin-like conditions.

\section{Self-stress of a Spherical Metallic Shell}
Classically a massless particle in a spherical cavity is an integrable system,
but the semiclassical spectrum is only asymptotically correct.  One therefore cannot expect the semiclassical approximation to
be exact in this case.  It nevertheless will be surprisingly accurate. The SCE is obtained by performing
the integrals of\equ{dualtrafo} in stationary phase and has a very transparent interpretation in terms of
periodic orbits \emph{within} the cavity only. The sign of the SCE of a spherical cavity in particular will be
quite trivially established and the discrepancy of 1\% with the field-theoretic results may very well largely be due to
diffractive corrections from creeping orbits that wind about the exterior of the sphere.
As argued above there are no (potentially divergent) local contribution
to the Casimir energy from such an idealized surface in the electromagnetic case -- its local surface tension in
fact vanishes\cite{BD04,BGH03}. The only subtraction in the spectral density required for a finite Casimir energy
with ideal metallic boundary conditions is the Weyl contribution proportional to the volume of the sphere.
This subtraction corresponds to ignoring the ${\bf m}=(0,0,0)$ term in the sum of\equ{dualtrafo}.  The
remaining difficulty in calculating the SCE is a convenient choice of action-angle
variables. For a massless scalar in three dimensions satisfying boundary conditions with spherical symmetry, an
obvious set of actions is the magnitude of angular momentum, $I_2=L$, one of the components of angular momentum
$I_3=L_z$ and an action $I_1$ associated with the radial degree of freedom.

Since the azimuthal angle of any classical orbit is constant, the energy $E=H(I_1,I_2)$ of a massless particle
in a spherical cavity of radius $R$ does not depend on $I_3=L_z$. In terms of this choice of actions,
the classical energy is implicitly given by,
\bel{H}
\pi I_1 + I_2 \arccos\left(\frac{c I_2}{E R}\right)=\frac{E R}{c}
\sqrt{1-\left(\frac{c I_2}{E R}\right)^2}\ .
\ee
The  branches of the square root and inverse cosine in \equ{H} are chosen so that $I_1$ is positive. It is
convenient to introduce dimensionless variables
\bel{var}
\lambda=2 E R/(\hbar c)\ \ {\rm and}\ \ z= c I_2/(E R),
\ee
for the total energy  (in units of $\hbar c/(2R)$) and the angular momentum (in units of $E R/c$) of an orbit.
Note that $z\in [0,1]$ and that the semiclassical regime formally corresponds to $\lambda\gg 1$, i.e. to
wavelengths that are much less than the dimensions of the cavity. Using\equ{H} and the definitions in \equ{var},
the angular frequency of the radial motion is $\omega^{-1}=(\partial E/\partial I_1)^{-1}=R\sqrt{1-(cI_2/(E
R))^2}/(\pi c)=(R/\pi c)\sqrt{1-z^2}$.

With the help of\equ{H} and the definitions of\equ{var}, the semiclassical expression in\equ{dualtrafo} for the
Casimir energy of a massless scalar field satisfying Neumann or Dirichlet boundary conditions on a spherical
surface becomes,
\bal{sphere}
{\cal E}^{\rm sph}&=&\frac{\hbar c}{4\pi R}{\sum_{n,w\geq 0}}^\prime
\Re\left[e^{-i\frac{\pi}{2}\beta(n,w)}\times\right.\nonumber\\
&&\hspace{-5em}\left.\times  \int_0^\infty
\hspace{-1em}d\lambda\lambda^3\hspace{-.5em}\int_0^1
\hspace{-.5em}dz z {\scriptstyle \sqrt{1-z^2}}\; e^{i\lambda[n(\sqrt{1-z^2}-z\arccos(z))+w\pi z]}\right]\
.\nonumber\\
\ea
The integral over $I_3$ has here been performed in stationary phase approximation. Because the Hamiltonian does
not depend on $I_3$, only periodic orbits with $m_3=0$ contribute\cite{Schaden06} significantly in stationary phase. Since
$-I_2\leq I_3\leq I_2$, one has that $\int dI_3= 2I_2=\lambda z$. The factor $2 I_2$ accounts for the
$2(l+1/2)$-degeneracy of a state with angular momentum $L=l+1/2=I_2$. By taking (4 times) the real part
in\equ{sphere} one can restrict the summations to non-negative integers and choose the positive branch of the
square root- and inverse cosine- functions in the exponent. The primed sum  here means that the summand is weighted by half
if one of the integers vanishes and the absence of the $n=w=0$ term. The Keller-Maslov index
$\beta(n,w)$ of a classical sector depends on whether Neumann or Dirichlet boundary conditions are satisfied on
the spherical shell and is obtained below.

For positive integers $w$ and $n$, the phase of the integrand in\equ{sphere} is stationary at
$z=\zb(n,w)\in[0,1]$ where,
\bal{sp}
0&=&-n\arccos(\zb)+w\pi\nonumber\\
&&\hspace{-2em}\Rightarrow \ \zb(n,w)=\cos(w\pi/n),\ n\ge 2 w>1\ .
\ea
Restrictions on the values of $w$ and $n$ arise because $\arccos(\zb)\in[0,\pi/2]$ on the chosen branch. The
phase is stationary at classically allowed points only for sectors with $n\ge 2 w>1$. Semiclassical
contributions to the integrals of other sectors arise due to the endpoints of the $z$-integration at $z=0$ and
$z=1$ only. These diffractive contributions are of sub-leading order in an asymptotic expansion of the
spectral density for large $\lambda$ and will be ignored. Note that $w\rightarrow w+n$ just amounts to choosing 
another branch of the inverse cosine.

The classical action in sectors with stationary points is,
\bal{Scl}
S_{cl}(n,w)&=&\hbar \lambda n
\sin(w\pi/n)\\
&&\hspace{-4em}=(E/c) 2 n R\sin(w\pi/n)=(E/c)L(n,w)\ ,\nonumber
\ea
where $L(n,w)$ is the total length of the classical orbit. Some of these classical periodic orbits are shown in
Fig.~1. The integer $w$ in\equ{Scl} gives the number of times an orbit circles or winds about the origin. The integer $n>1$
gives the number of reflections off the spherical surface (windings of the radial motion). As indicated in Fig.~1, the envelope of  the set of classical periodic orbits in the $(n,w)$-sector is a caustic surface and a double covering is required for a unique
phase-space description\cite{Keller58}. The two sheets are joined at the inner caustic [indicated by a dashed
circle in Fig.~1] and at the outer spherical shell of radius $R$. Every orbit that passes the spherical shell
$n$ times also passes the caustic $n$ times. The cross-section of a bundle of rays is reduced to a point
at the spherical caustic surface. The caustic thus is of second order and associated with a phase loss of $\pi$
every time it is crossed. At each specular reflection off the outer shell Dirichlet boundary conditions require
an additional phase loss of $\pi$  whereas there is no phase change for Neumann boundary conditions. Altogether
the Keller-Maslov index of the sector $(n,w)$ depends on $n$ only and is given by,
\bel{KM}
\beta(n,w)=\left\{\begin{array}{rl} 0, &\mathrm{for~Dirichlet~b.c.}\\
2n, &\mathrm{for~Neumann~b.c.}\end{array}\right. \ .
\ee
As noted in sect.~2, the electromagnetic field satisfying (ideal) metallic boundary conditions on a spherical shell,
may be viewed as two massless scalar fields of vanishing conformal dimension, one satisfying Dirichlet and
the other Neumann boundary conditions\cite{BD04}. Due to the Keller-Maslov phases of \equ{KM} only sectors
$(n,w)$ with \emph{even} $n=2 k\ge 2 w\ge 1$ contribute\cite{BD04} to the SCE in the electromagnetic case.

\begin{figure}[hpbt]
\includegraphics[width=3.5in]{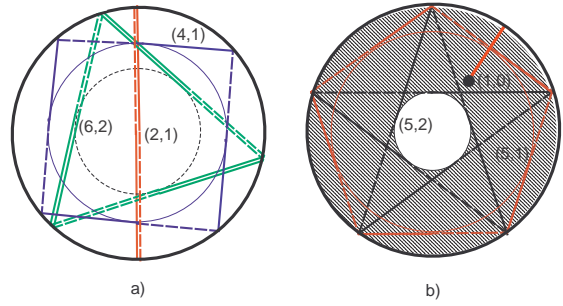}
\label{paths}
\caption{\small (Color online) Classical periodic rays of a spherical and
cylindrical cavity. a) The shortest primitive rays corresponding to sectors $(n,w)\in\{(2,1),(4,1),(6,2)\}$ that contribute to the electromagnetic SCE.
b) closed paths in sectors $(n,w)\in\{(1,0),(3,1),(5,1),(5,2)\}$ that reflect an odd number of times off the surface and whose contribution to the electromagnetic SCE vanishes. Caustic surfaces are indicated as thin circles. The
dashed part of any trajectory is on one sheet and its solid part on the other of a two-sheeted covering space.
The "phase space" of the $(5,2)$ sector is indicated by the hatched area. Note that the caustics are of $2^{{\rm nd}}$
order for a spherical cavity but of $1^{{\rm st}}$ order for a cylindrical one.}
\end{figure}

The classical action vanishes for sectors with $w=0$ or $n=0$ and these sectors do not contribute to the SCE in leading approximation. The $n=0$ sector corresponds to paths that do not reflect off the shell and thus gives rise to volume contributions that are subtracted.   \equ{sp}
implies that extremal paths in the $(n>0,w=0)$ sectors have maximal angular momentum $1=\zb=l c/(E R)$. These
are great circles that lie wholly within the spherical shell. These paths are not stationary and lead to diffractive contributions that we will ignore here.  For $(n>0,w>0)$ the curvature of the exponent at $\zb(n,w)$ is finite,
\bel{Sdd}
\left.\frac{\partial^2}{\partial z^2}[n({\scriptstyle
\sqrt{1-z^2}}-z\arccos(z))+w\pi z]\right|_{\zb(n,w)}=\frac{n}{\sin({\frac{w\pi} n})}\ ,
\ee
whereas it diverges in sectors with $w=0$. The behavior of the exponent for $z\sim 1$ in this case is,
\bel{m0}
{\scriptstyle \sqrt{1-z^2}}-z\arccos(z)={\scriptstyle \frac{2\sqrt{2}}{3}(1-z)^{3/2}}+ {\cal O}({\scriptstyle
(1-z)^{5/2}})\ .
\ee
Quadratic fluctuations about the classical orbit with $w=0$ thus are of vanishing width and these sectors do not
contribute in stationary phase approximation. To leading semiclassical accuracy, the Casimir energy of a
spherical cavity with an ideal metallic boundary therefore is,
\bal{sphere1}
{\cal E}^{\rm sph}_{\rm EM}&\sim&\frac{\hbar c}{4\pi R}\,\Re\sum_{n=1}^\infty
(1^n+(-1)^n)\sum_{w=1}^{n/2} \nonumber\\
&&\hspace{-3.5em}\times  \int_0^\infty \hspace{-1em}d\lambda\lambda^3e^{i n \lambda\sin({\frac{w\pi}n})}\hspace{-.5em}\int_0^1 \hspace{-.5em}dz
z{\scriptstyle \sqrt{1-z^2}} \; e^{i \frac{n\lambda(z-\zb(n,w))^2}{2\sin(w\pi/n)}}\nonumber\\
&&\hspace{-3.5em}\sim\frac{\hbar c}{R}\left[\sum_{n=1}^\infty \frac{1}{16\pi n^4} +\sum_{n=2}^\infty\frac{15
\sqrt{2}}{256 n^4} \sum_{w=1}^{n-1}\frac{\cos(\frac{w\pi}{2n})}{
\sin^2(\frac{w\pi}{2n})}\right]\nonumber\\
&&\hspace{-3.5em}\sim 0.04668...\frac{\hbar c}{R}\ .
\ea

This semiclassical estimate is only about 1\% larger than the best numerical value\cite{Milton78}
$0.04617...\hbar c/R$ for the electromagnetic Casimir energy of a spherical cavity with an infinitesimally thin
metallic surface. Note that the contribution from the $(2w,w)$ sectors had to be considered separately
in\equ{sphere1} since the measure $dz z$ vanishes at the stationary point $\zb(2w,w)=\cos(\pi/2)=0$ of the
integrand, which is an endpoint of the integration domain. As can be seen in Fig.~1a), the classical rays of
$(2w,w)$-sectors go back and forth between antipodes of the cavity and pass through its center -- they have
angular momentum $\vec L=0$.

The shortest primitive orbits give somewhat less than half [$1/(16\pi)\sim 0.02$] of the total SCE of the
spherical cavity -- much less than the 92\% they contribute to the Casimir energy of parallel plates. The
reason is that contributions only drop off as $1/n^2$ rather than like $1/n^4$ as for parallel plates. The
length of an orbit in the $(4,1)$-sector (the inscribed square in Fig.~1a) furthermore is just a factor of
$\sqrt{2}$ longer than an $(2,1)$-orbit [which in turn is a factor of $1/\sqrt{2}$ shorter than an
$(4,2)$-orbit]. To estimate the magnitude of the contribution of any particular sector one has to take the
available phase space as well as the ray's length into account. Thus, although the length of a $(2 n,1)$-orbit
tends to $2\pi R$ for $n\rightarrow \infty$, the associated phase-space (essentially given by the volume of the
shell between the boundary of the cavity and the inner caustic) decreases like $1/n^2$.  This accounts for the
relatively slow convergence of the sum in\equ{sphere1}. To achieve an accuracy of $10^{-5}$, the first 50 terms
of the sum were evaluated explicitly  and the remaining contribution was estimated using Richardson's
extrapolation method. However, note that the semiclassical 2-reflection coefficient is just
$2/5$-ths of the leading field-theoretic coefficient for a dilute dielectric-diamagnetic spherical shell\cite{Klich99}.
We will encounter a similar discrepancy in the case of a cylindrical shell and will discuss it further in the conclusion.

\section{Self-stress of a Cylindrical Metallic Shell}
The example of a spherical cavity shows that one may obtain electromagnetic Casimir self-energies rather accurately by
considering only small fluctuations about their classical periodic orbits. Since the classical periodic rays of finite
length for a long cylindrical shell are the same as for a sphere, one would expect that a semiclassical calculation of
the self-stress  is just as straightforward for a cylindrical cavity.

This is not the case.  We verify below that the electromagnetic SCE of an ideal metallic cylindrical shell vanishes in stationary phase approximation\cite{MSSS03,Schaden06b}. The contribution from any periodic orbit to the electromagnetic SCE vanishes in this approximation for
the same reason that it is positive for a spherical cavity -- due to optical phases. However, going beyond stationary phase approximation and including certain Fresnel diffraction effects, we again obtain a very good approximation to the Casimir self-stress of a metallic cylinder from quadratic fluctuations about classical periodic paths.

The cylinder appears particularly suited for a semiclassical analysis in terms of massless scalars
because the transverse electric- and magnetic- modes satisfy Dirichlet and Neumann boundary conditions
on the cylindrical surface. The classical system is integrable and we could directly employ the formalism of Berry and Tabor\cite{BT76,Schaden06} or an extended Gutzwiller approach\cite{Brackbook,MSSS03} to obtain the SCE in much the same way as we did for the sphere. However, to better understand why the SCE vanishes in this approximation and improve upon it, we first obtain the dual expression for the Casimir self-energy  of a cylindrical cavity directly from the semiclassical (WKB) estimate of the eigenvalues of the scalar fields. The Casimir energy of an  infinitesimally thin metallic cylindrical shell of radius $R$ within a much larger cylinder of fixed radius $R_>\sim\infty$ in principle is given by the $R$-dependent part of the zero-point energy,
\bal{CasCyl}
{\cal E}^{\rm cyl}_{EM}(R)&=&\lim_{R_>\rightarrow\infty}\frac{\hbar c L}{2\pi}\sum_{{\genfrac{}{}{0pt}{}{D,N}{n}}}\int^\infty_0 \hspace{-.5em}dq\left\{\sqrt{q^2+\kappa^2_n(0,R)}\right.\nonumber\\
&&\hspace{-3em}\left.+\sqrt{q^2+\kappa^2_n(R,R_>)}-\sqrt{q^2+\kappa^2_n(0,R_>)}\right\}_{D,N},
\ea
where $\{\kappa_n(R_<,R_>)\}_{D,N}$ is the spectrum of wave numbers for a scalar field satisfying Dirichlet$(D)$ or Neumann$(N)$ boundary conditions on a 2-dimensional annulus with inner radius $R_<$ and outer radius $R_>$.

The asymptotic heat-kernel expansion\cite{Kirsten02} implies that the subtracted expression in \equ{CasCyl} is finite: the potentially logarithmic divergence proportional to the average of the 3rd power of the (extrinsic) curvature of the cylindrical surface is canceled. Milton and DeRaad calculated this finite self-energy exactly some time ago\cite{DRM81}. Since they uniformly approach the exact wave numbers sufficiently rapidly, the SCE obtained by replacing the exact eigenvalues in\equ{CasCyl} by their WKB-estimates also is finite. For $R_>\sim\infty$ all periodic orbits in the annulus have a length of ${\cal O}(R_>)$ and the only finite contribution to the SCE from the annulus is due to (diffractive) creeping orbits that wrap about the inner cylinder. We will neglect this small contribution to the SCE and consider only radius-dependent semiclassical contributions from the inner cylinder. The transverse wave numbers $\kappa_{n\ell}(0,R)=x_{n\ell}/R$  of \emph{interior} modes of the cylinder in WKB-approximation are positive solutions of\cite{Brackbook,KR60},
\bel{EVS}
f_\ell(x_{n\ell})=\pi(n+\half\pm\quarter), \text{ for } n=0,1,\dots; \ell=0,1,\dots\ ,
\ee
where the $(+)$- and $(-)$-sign corresponds to Dirichlet and Neumann boundary conditions respectively and,
\bel{fdef}
f_\ell(x)=\sqrt{x^2-\ell^2}-\ell\arccos(\ell/x)\ .
\ee
This semiclassical (Debye) approximation generally gives the zeros of Bessel functions of the first kind and their derivatives to better than 1\%. There is but one notable exception: the zero of $J'_0(x)=J_1(x)$ at $x=0$ corresponds to a WKB value of $x_{00}=\pi/4$. Note that $f_\ell(x=\ell)=0$ and all semiclassical wave numbers satisfy $x_{n\ell}>\ell$.

Inserting the approximation in \equ{EVS} and \equ{fdef} for the eigenvalues, the $R$-dependent contribution of interior modes to the SCE of a conducting cylindrical shell is the finite $R$-dependent part of,
\bal{CasCylSC}
&&\hspace{-2em}{\cal E}^{\rm cyl}_{\rm EM}\!\sim\!\frac{\hbar c L}{2\pi^2 R^2}\hspace{-.5em}\sum^\infty_{n=-\infty} \hspace{-.5em}(i^n+(-i)^n) \hspace{-.3em}\sum_{\ell=0}^\infty\nonumber\\
&&\times\int_0^\infty\hspace{-1.2em} dy\!\int_{\ell}^\infty\hspace{-1em}dx\sqrt{y^2\!\!+\!\!x^2}f'_\ell(x) e^{2 i n f_\ell(x)}\nonumber\\
&&\hspace{-2em}\sim\!-\frac{\hbar c L}{2\pi^2 R^2}\Im\sum_{\ell=0}^\infty\int_0^\infty\hspace{-1.2em} dy\!\int_{\ell}^\infty\hspace{-1em}dx\sqrt{y^2\!\!+\!\!x^2} \partial_x\ln\left[\cos(2f_\ell(x))\right]\ .\nonumber\\
\ea
The summation over $n$ has been performed in the last expression. It shows the equivalence of the present approach with one based on the generalized argument principle\cite{VanKampen} with the function, $\cos(2f_\ell(x))=\half(e^{-i f_\ell(x)}+i e^{i f_\ell(x)})(e^{-i f_\ell(x)}-i e^{i f_\ell(x)})$, whose zeros are the WKB estimates for the mode frequencies for each partial wave, and a contour that runs from $x=\infty$ to $x=\ell$ just below the real axis and returns to $x=\infty$ just above it. Although divergent because it includes only the contribution from the interior, it is reassuring that the expressions in\equ{CasCylSC} are not logarithmically divergent. The divergences therefore can be subtracted unambiguously (and in fact are canceled by exterior contributions we are not considering) and a semiclassical evaluation is possible.

One arrives at \equ{CasCylSC} by applying Poisson's resummation formula,
\bel{Poisson}
\sum_{n=-\infty}^{\infty} \delta(f-n) =\sum_{n=-\infty}^\infty e^{2\pi i n f}\ ,
\ee
to the dimensionless (and scaled) semiclassical spectral densities,
\bal{rho}
\rho_{D/N}(x)&=&\sum_{\ell=0}^\infty\sum_{n=0}^\infty \delta(x-x_{n\ell})\\
&&\hspace{-4em}\sim\sum_{\ell=0}^\infty\theta(x-\ell)f'_\ell(x)\sum_{n=-\infty}^\infty \delta(f_\ell(x)-(n+\half\pm \quarter)\pi)\ .\nonumber
\ea
The Heaviside function $\theta(x-\ell)$ in \equ{rho} ensures that $f_\ell$ is real.  Using Poisson's relation once more in the form,
\bel{Poisson0}
\sum_{\ell=0}^{\infty} g(\ell)=\half g(0)+\sum_{w=-\infty}^\infty \int_0^\infty d\ell\; e^{2\pi i w \ell}g(\ell)\ ,
\ee
the sum over partial waves of cavity modes is converted to one over the winding number of classical paths.
In terms of the dimensionless wave number $\lambda=\sqrt{x^2+y^2}=E R/(\hbar c)$ and the longitudinal- and angular- momentum fractions $\alpha=y/\lambda$ and $z=\ell/x$ one obtains,
\bal{CasCylSC1}
{\cal E}^{\rm cyl}_{\rm EM}\sim\frac{\hbar c L}{\pi^2 R^2}\hspace{-.5em}\sum^\infty_{n=-\infty}\hspace{-.5em} (-1)^n \hspace{-.5em}\sum_{w=-\infty}^\infty\int_{0}^\infty\hspace{-1em} \lambda^3 d\lambda\int_0^{1}\hspace{-.5em}d\alpha\hspace{2em}&&\\
\times \int_0^1\hspace{-1em} \sqrt{1-z^2}e^{2 i \lambda\sqrt{1-\alpha^2}(4 n (\sqrt{1-z^2}-z\arccos{z})+w\pi z) }dz\ .\nonumber&&
\ea
We here used that only terms with even $n$ contribute in\equ{CasCylSC}. Only the principal branch of the inverse cosine is to be considered here and the integrals are formal in the sense that they are to be evaluated to leading (non-vanishing) order in stationary phase approximation only. The integrals are finite in this restricted sense and their divergent part has been implicitly subtracted. They otherwise diverge (in all sectors) due to the behavior of the integrand at large $\lambda$ near $\alpha\sim 1$, i.e. for large longitudinal momentum fractions. As explained in sect.~2 we are assured that all surface divergences cancel for metallic boundary conditions on a cylinder. We are interested only in the finite contributions arising from quadratic fluctuations about periodic classical trajectories (stationary points of the integrand in the $(n\neq 0,w\neq 0)$ sectors).

The $\half g(0)$ term in \equ{Poisson0} gives a surface correction\cite{Brackbook} to \equ{CasCylSC1}. As mentioned in sec.~2, such surface contributions cancel in the electromagnetic case. One may explicitly verify this by subtracting the $n=0$ sector and noting that
\bal{0corr}
&&\lim_{\eps\rightarrow 0^+}\Re\sum^\infty_{n=1} (-1)^n\int_0^\infty\hspace{-1em} d\lambda\; \lambda^2 \int_0^{\pi/2-\eps}\hspace{-1.5em}d\phi\; e^{4 i n \lambda \cos{\phi} }\nonumber\\
&&=\frac{3\zeta(3)}{128}\lim_{\eps\rightarrow 0^+}\Re\ \left(i \int_0^{\pi/2-\eps}\frac{d\phi}{\cos^3{\phi}}\right)=0\ .
\ea

The stationary points of the $z$-integral in\equ{CasCylSC1} for $w\leq n>0$ at $\bar z_{nw}=\cos{(\pi w/n)}$  are the same as for a spherical cavity and correspond to planar periodic orbits like those of Fig.~1a. To quadratic order the fluctuations about the stationary point at $\alpha=0$ give rise to  Fresnel-like integrals of the type (for generic $\gamma>0$),
\bal{Fresnel}
\int_0^1 \hspace{0em}d\alpha\; e^{i\pi\gamma^2\sqrt{1-\alpha^2}}&\sim & e^{i\pi\gamma^2}\int_0^1 \hspace{0em}d\alpha\; e^{-i\pi\gamma^2\alpha^2/2}=\ \ \\
&&\hspace{-4em}= e^{i\pi\gamma^2}\left(C(\gamma)-i S(\gamma)\right)/\gamma\ .\nonumber
\ea
Extending the upper bound of the fluctuation integral in \equ{Fresnel} to $\infty$ and thereby replacing the cosine- and sine- Fresnel integrals of\equ{Fresnel} by their mean value of $1/2$ gives a vanishing value for the SCE of a metallic cylindrical shell\cite{MSSS03, Schaden06b}. The leading non-vanishing contribution in this case arises from the finite upper bound of the fluctuation integral in\equ{Fresnel}. The SCE of a metallic cylindrical shell in this semiclassical sense is almost entirely due to Fresnel diffraction effects.

The semiclassical evaluation of the integrals in \equ{CasCylSC1} again is facilitated by noting that one can take (4 times) the real part of the $(n>0,w>0)$-sector contributions and that the spectral density is analytic in the first quadrant. The $\lambda=i\xi$ integral in\equ{CasCylSC1} may thus equally well be performed along the positive imaginary axis. The contour is closed in the complex $\lambda$-plane by a large quarter circle on which the integral, or better, a regularized version of it, vanishes sufficiently rapidly. The integrand is real on the imaginary energy axis. For $0<w\leq n$ the stationary points of the integrand in \equ{CasCylSC1} are at $z=\cos\frac{\pi w}{2 n}$ and $\alpha=0$. They correspond to planar periodic rays of a cylindrical cavity like those in Fig.~1a with winding number $w$ and $2 n$ vertices. Expanding to quadratic order about these stationary points, we first integrate out fluctuations in $z$. Rescaling  $\xi\rightarrow\xi/(1-\alpha^2/2)$ the remaining $\xi-$ and $\alpha-$integrations may then be carried out independently. Note that the analytic continuation of the integrals is possible only if we assume an upper bound at $\alpha=1-\eps$.  Starting with the expression of \equ{CasCylSC2} this procedure gives the following semiclassical estimate of ${\cal E}^{\rm cyl}_{\rm D/N}$ (with $c_{nw}:=\cos{\left(\frac{\pi w}{2 n}\right)}$ and $s_{nw}:=\sin{\left(\frac{\pi w}{2 n}\right)}$)
\bal{CasCylSC2}
{\cal E}^{\rm cyl}_{\rm EM}&\sim&\frac{4\hbar c L}{\pi^2 R^2}\sum_{n=1}^\infty(-1)^n\sum^{n}_{w=1} \int_{0}^\infty\xi^3 d\xi\int_0^{1}d\alpha\nonumber\\
&&\hspace{-4em}\times \int_0^1\hspace{-.5em} dz s_{nw}\exp\left[-4 n\xi s_{nw}\Big(1+\frac{(z-c_{nw})^2}{2 s^2_{nw}}-\frac{\alpha^2}{2}\Big)\right]\nonumber\\
&\sim & \frac{\hbar c L 15\sqrt{2}}{R^2 512\pi }\sum^{\infty}_{n=1} \frac{(-1)^n}{n^4}\left(-\half+\sum_{w=1}^n s_{nw}^{-2}\right)\nonumber\\
&&\hspace{2em}\times\ \int_0^{1}d\alpha (1-\frac{\alpha^2}{2})^{-7/2}\\
&=& \frac{7\pi (7\pi^2-240)}{276480}\frac{\hbar c L}{ R^2}= -0.013594\dots\frac{\hbar c L}{ R^2}\ .\nonumber
\ea

The SCE of a conducting cylindrical shell thus differs by less than 0.25\% from the exact field theoretic value\cite{DRM81} of $-0.0135613\dots\hbar c L/R^2$ if a certain kind of Fresnel diffraction is included. It was necessary to go beyond the formally leading semiclassical approximation because the latter violates causality and vanishes. Enforcing the bound on longitudinal momenta,  the semiclassical approximation again closely reproduces the Casimir self-stress on a metallic cylindrical shell.

\begin{figure}[hpbt]
\includegraphics[width=3.5in]{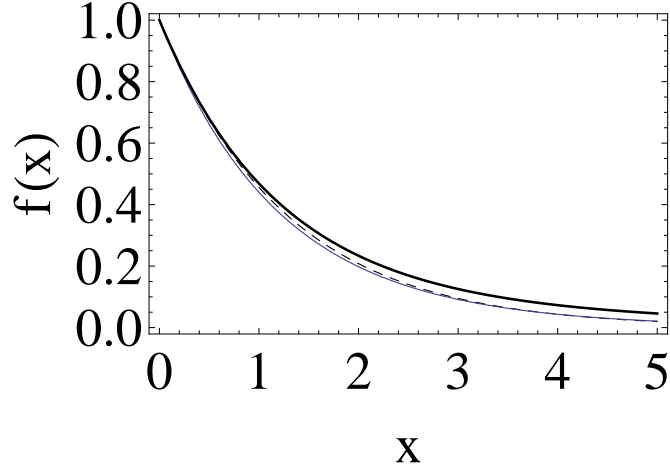}
\label{compare}
\caption{\small Integrating over the longitudinal momentum fraction. The upper solid curve shows the exact integral in\equ{integrals} over the longitudinal momentum fraction $\alpha$, the lower solid line is its semi-classical approximation to quadratic order in the fluctuations. The dashed curve gives the simple exponential approximation $e^{-\pi x/4}$.}
\end{figure}

To better understand the approximation made, compare the following approximations to the Fresnel-like $\alpha$-integral in\equ{CasCylSC1}:
\bal{integrals}
&&\int_0^1 \hspace{-.5em}d\alpha\, e^{-x\sqrt{1-a^2}}=\frac{\pi}{2}\left( H_{-1}(i x)-I_1(x)\right)\sim e^{-\pi x/4}\nonumber\\
&&\hspace{1em}\sim \int_0^1\hspace{-.5em} d\alpha\, e^{-x(1-\frac{a^2}{2})}=-i e^{-x}\sqrt{\frac{\pi}{2 x}}\text{erf}\left(i\sqrt{\frac{x}{2}}\right)
\ea
The first line expresses the $\alpha$-integral occurring in\equ{CasCylSC1} for values of the energy on the positive imaginary axis in terms of Struve- and Bessel- functions. The function $e^{-\pi x/4}$ gives the best exponential fit for small $x\sim 0$. The second line of\equ{integrals} is the semiclassical approximation to second order in the fluctuations. It also  gives a uniform approximation to the integral that decays exponentially for $x\sim\infty$ (but as $e^(-x/2)/x$). As shown in Fig.~2, both approximations reproduce the integral well for small values of $x$ -- where the result is sizable -- but cut down its slow decay ($\sim 1/x^2$) at large $x\sim\infty$. As noted before, the power-like decay of the original integral gives rise to divergent contributions to the Casimir energy in every $(n,w)$ sector\footnote{Note that the power-like decay arises from the integration-region $\alpha\sim 1$ where the longitudinal wave number is much larger than the transverse one. It may be regularized by imposing $\alpha<1-\eps$ with finite $\eps>0$.}. Assuming (and heat-kernel considerations show this is possible for an ideal metallic shell\cite{BKV99,BP01}) all divergences are subtracted unambiguously, the leading non-vanishing semi-classical approximation apparently is quite a reasonable estimate of the finite part. One in principle could improve on the representation of the integral by including higher orders in the expansion of $\sqrt{1-\alpha^2}$ about $\alpha=0$ -- but these contributions would be inconsistent with the (leading order) WKB approximation to the eigenvalues.  We can to some extent assess the sensitivity of the SCE to the precise manner in which the power-like tail is cut off by comparing with the exponential fit $e^(-\pi x/4)$ for the $\alpha$-integral: in this approximation the Casimir energy for a metallic cylinder becomes $(7\pi^2-240)\hbar c L/(288\pi^3\sqrt{2}R^2)=-0.013533\dots \hbar c L/R^2$ -- well within the error of the semiclassical estimate.

\section{Discussion and Conclusions}
We here obtained the electromagnetic Casimir self-stress of perfectly conducting spherical and cylindrical shells in semiclassical
approximation. This approach reproduces the field theoretic values\cite{Boyer68, Milton78} to better than 1\%. The semiclassical description by two massless scalars in general gives the \emph{electromagnetic} Casimir-stress of a  metallic spherical shell rather than the sum of the Casimir stresses due to scalar fields satisfying Neumann and Dirichlet boundary conditions. For a spherical shell the latter has the opposite sign and is an order of magnitude larger\cite{BM94,NP97,GR98}. The semiclassical description is inherently conformal and thus may be best suited to describe scalar fields that satisfy conformally covariant boundary conditions\cite{Gilkey}. It was previously observed\cite{Schaden06} that the dispersion for massless scalar particles leads to a semiclassical description of conformally coupled scalar fields on curved spaces such as $S_3$. They also appear to couple conformally to curved boundaries. "Neumann" boundary conditions are semiclassically imposed by requiring no phase change and specular reflection. They do not change under conformal rescaling of the boundary and in this sense are conformally covariant. For spherical boundaries they describe (massless) scalar fields satisfying conformally covariant\cite{Gilkey} Robin conditions. This conjecture is supported by the fact that the low-lying semiclassical eigenfrequencies implied\footnote{one can extract the semiclassical estimate of the eigenfrequencies from the spectral density of\equ{sphere1} by reversing the procedure used to obtain the semiclassical spectral density in \equ{CasCylSC} from the WKB-estimate of the eigenfrequencies in the case of a cylindrical shell.} by \equ{sphere1} are closer to those of two scalar fields satisfying Dirichlet- and Robin- than Dirichlet- and Neumann- conditions.  Robin boundary conditions approach Neumann boundary conditions at high frequencies and both would be similarly implemented in geometrical optics (i.e. with specular reflection and no phase loss).

On a cylindrical surface in flat space ordinary Neumann and Dirichlet conditions are already conformally covariant. We estimated the electromagnetic Casimir self-stress of a metallic cylindrical shell by explicitly summing WKB-approximations to the eigenfrequencies of the two scalars in the dual picture using Poisson's resummation formulae. The dual of the principal quantum number gives the number of interactions with the shell. Summing over these one arrives at the general argument principle (see \equ{CasCyl}) often used\cite{VanKampen} to evaluate Casimir energies when the spectrum is only implicitly known. The difference here is that the roots of the characteristic function are WKB estimates of the eigenfrequencies. The dual to the sum over partial waves is the sum over windings about the origin. The stationary points of minimal action in each sector are classical periodic trajectories. We evaluate the integrals in each non-trivial sector to quadratic order in the fluctuations\cite{BT76,Brackbook}. For a cylindrical shell it is essential to enforce the upper bound on the longitudinal momentum fraction $0\leq\alpha<\leq 1$.
Ignoring it and integrating without restriction over quadratic fluctuations about the stationary point at $\alpha=0$, the SCE of a cylindrical shell in fact vanishes\cite{MSSS03,Schaden06}.  However,  restricting the quadratic fluctuations to $\leq 1$ essentially reproduces the field theoretic value for the Casimir energy of a metallic cylinder. The Casimir energy of a cylindrical metallic shell in a semi-classical sense is entirely due to Fresnel diffraction effects.  The restriction of fluctuations to $\alpha<1$ in fact is required by causality: the spectral density otherwise is not analytic in the first quadrant. This violation of causality does not occur for any other fluctuation integral, which to leading order are evaluated without restriction.

From a semiclassical point of view, the metallic cylindrical shell thus is a rather interesting geomery and conceptually more rewarding than the spherical one. Because a straightforward semi-classical evaluation to leading order resulted in a vanishing Casimir stress, it was previously believed\cite{Schaden06b} that this explained the \emph{exact} vanishing of the Casimir stress on a dielectric-diamagnetic cylindrical shell (with equal speed of light on either side) to first order in the reflection coefficients\cite{BD78,MNN99,NP98}.  The non-vanishing Casimir energy of a metallic cylindrical shell was attributed to sub-leading diffraction effects. The last conjecture has now been verified, but the diffractive contribution due to classical orbits with just two reflections, the $n=1$ contribution\footnote{The factor $\left(-\half+\sum_{w=1}^n s_{nw}^{-2}\right)=(4 n^2-1)/6$ in \equ{CasCylSC2} vanishes for $n=1/2$, not $n=1$!} in\equ{CasCylSC2}, is $-0.0174076\dots \hbar cL/R^2$. It does not vanish and is the largest contribution in magnitude, larger than the total self-stress of the metallic cylinder. Although we thus lack a semiclassical understanding of a weakly reflecting dielectric-diamagnetic cylindrical shell, it should be pointed out that the self-stress of a dielectric-diamagnetic spherical shell\cite{Klich99} also is underestimated by a factor of $2.5$ in this approximation.

In the limit of very small reflection coefficients it probably is impossible to ignore contributions from the exterior. It may be necessary to include diffractive contributions from paths that creep about the cylinder to semiclassically describe weakly reflective interfaces. Mathematically the exact cancelation in the cylindrical case is an addition theorem for Bessel functions that explicitly requires the exterior contributions\cite{PMK07}. That perturbation in the reflection coefficients is quite delicate becomes evident for dielectric cylinders\cite{RM05} and spheres\cite{BMM99}: to second order in the reflection coefficients the results for the Casimir stress in this case are finite and comparable to the case where the speed of light is continuous across the boundary, but the total Casimir self-stress diverges logarithmically\cite{BKV99,BP01} if the speed of light in the interior and exterior do not match exactly.

{\bf Acknowledgements:} I greatly enjoyed the hospitality of Texas A\&M and New York University during part of this work and wish to especially thank S. Fulling, D. Zwanziger, K.A. Milton, M. Bordag, I. Klich and K. Shajesh for illuminating discussions and encouragement.   This work is partially supported by the National Science Foundation with Grant No.~PHY0902054.

\end{document}